\documentclass{ws-procs9x6}
\def\delslash{\,\,{\raise.15ex\hbox{/}\mkern-10mu \partial}}
\newcommand{\lapprox}{\raisebox{-0.5ex}{$\
\stackrel{\textstyle<}{\textstyle\sim}\ $}}

\begin{document}

\title{Vector meson at non-zero baryon density and zero sound}

\author{S.J. Hands}

\address{Department of Physics, 
University of Wales Swansea, \\ 
Singleton Park, Swansea SA2 8PP, U.K.}

\author{C.G. STROUTHOS}
\address{Division of Science and Engineering, \\
Frederick Institute of Technology, Nicosia 1036, Cyprus}


\maketitle

\abstracts{We present simulation results of the $(2+1)d$ four-fermion model with a baryon chemical potential $\mu$.
We examine temporal correlation functions of the vector meson, 
and find evidence of phonon-like behavior characterised by a linear dispersion relation in the long 
wavelength limit. We also discuss the consistency of our numerical results with analytical solutions 
to the Boltzmann equation corresponding to zero sound. We argue that our results provide the first evidence 
for a collective excitation in a lattice simulation.
}


We study the $(2+1)d$ Gross-Neveu (GN) model with a $Z_2$ chiral symmetry 
describing $N_f$ flavors of self-interacting 
fermions, with continuum lagrangian 
\begin{equation}
\mathcal{L}=\bar\psi(\partial{\!\!\!/\,}+\mu\gamma_0)\psi
-{g^2\over{2N_f}}(\bar\psi\psi)^2.
\end{equation}
As $\mu$ is increased, the model exhibits a sharp first-order transition
at a critical $\mu_c$
from a phase where a fermion mass $m_f$ is dynamically generated but baryon density
$n_B=\langle\bar\psi\gamma_0\psi\rangle$ vanishes to one where $m_f=0$
and $n_B\propto\mu^2$ [1]. The main new feature emerging for $\mu>\mu_c$
is the existence of a new scale, the Fermi momentum $k_F$. Since the lowest energy excitations
of the ground state have $|\vec{k}| \approx k_F$, measurements of Euclidean timeslice
correlators with $\vec{k}\neq 0$ are mandatory. 

If certain conditions are fulfilled, such as an effective interaction between 
quasiparticles which is both short-ranged and repulsive, then there is a massless bosonic excitation 
in the spectrum as $T\to0$. Carrying zero baryon number, it is a {\it phonon}, 
ie. a quantum of a collective excitation called {\it zero} {\it sound} [2]. 
Sound propagation occurs in any elastic medium; zero sound happens when the elasticity 
originates not from collisions between individual particles, but from the force
on a single particle due to its coherent interaction with all others present in the 
medium. It can be pictured as a propagating distortion in the local shape of the Fermi
surface, and its speed of propagation exceeds that of conventional ``first'' sound.
In [3] we solved the self-consistent equation for zero sound in the framework of
Fermi liquid theory in the case of the $3d$ GN model to leading non-trivial order 
in $1/N_f$. We found solutions for sound propagation with speed $\beta_0 > \beta_F$, the Fermi velocity.
Here we discuss our numerical results and compare them with the 
analytical solutions presented in detail in [3]. 

The lattice regularized action of our model uses staggered fermions.
Details of the formulation and simulation algorithm are given
in [1,4].
The simulation parameters are $N_f=4$ and $a/g^2=0.75$, corresponding to a physical fermion mass at $\mu=0$
of $m_f\simeq0.17a^{-1}$. 
All results are taken in the chirally-restored phase,
ie. with $\mu>\mu_c\simeq0.16a^{-1}$.

By analysing the decay of the appropriate correlation function with Euclidean
time,
we calculated the dispersion relations $E(k)$
for the spin-${1\over2}$
quasiparticle, which carries a baryon charge, and
$\omega(k)$ for various meson states
of the form $\bar\psi\Gamma\psi$ which are best thought of as
excitations formed from a particle-hole
pair. We simulated $L_s^2\times48$ systems with $L_s=32$ and 48,
with $\mu$ ranging from 0.2 to 0.6,
and measured $E(k),\omega(k)$ for $\vec k=(k,0)$ with $k=0,2\pi/L_s,\ldots,\pi/2$.
The increased momentum resolution offered by $L_s=48$ has proved to be important. 
Note that the staggered fermion action is only invariant under translations
of two lattice spacings, which restricts the space of accessible momenta.
In [3,5] we measured the quasiparticle propagator and showed that the
transition between particle and hole behavior is rather sudden, characteristic of a
well-defined Fermi surface. We also presented [5] the theoretical predictions
of Fermi liquid properties to leading non-trivial order in $1/N_f$ for $\mu \gg \mu_c$. 
The results for the Fermi velocity $\beta_F$ and the Fermi momentum $k_F$ are given by:
\begin{equation}
\beta_F=1-{1\over{16N_f}}\simeq0.984\;\;\;;\;\;\;
{k_F\over{\mu\beta_F}}=1-{1\over{8N_f}}\simeq0.969.
\label{eq:fl1N}
\end{equation}
As discussed in [3,5] the agreement between the analytical predictions and the numerical 
results on $48^3$ lattices for $\mu=0.2,...,0.6$ and with discretization effects taken into account
is at best qualitative. It should be noted that analytic predictions of $\beta_F$ and $k_F$
to $O(1/N_f)$ for finite $\mu$ strictly requires a knowledge of the gap equation to two loops, 
which is not yet available.
For the remaining analysis we
will therefore assume the free field
values $k_F=\mu$, $\beta_F=1$.

\begin{figure}[t]
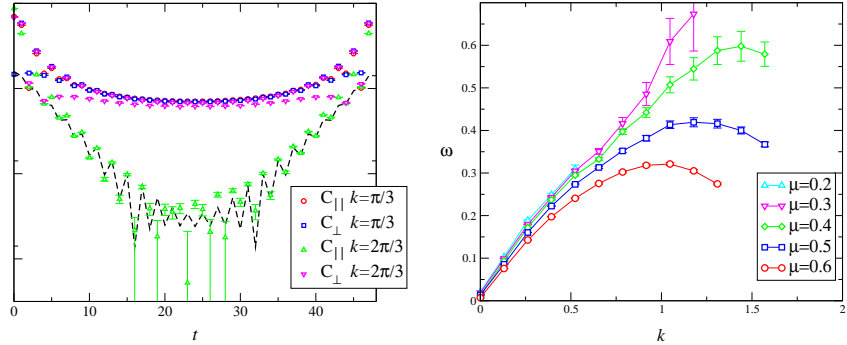

\begin{tabular}{ll}
{\includegraphics[scale=0.27]{fig4.eps}}&
{\includegraphics[scale=0.27]{fig5.eps}}
\end{tabular}
\caption{Left: Meson propagators $\vert C_\parallel\vert$ and $\vert C_\perp\vert$
for both $k={\pi\over3}$ and $\pi-{\pi\over3}$ for $\mu=0.5$. The dashed line
shows a fit of the form (\ref{eq:zsfit}). Right: Dispersion relation $\omega(k)$ for various $\mu$ for the
channel defined by the correlator
$C_\parallel(\pi-k)$. Data are from $48^3$. }
\label{fig:results}
\end{figure}

Next we consider the meson sector, ie. correlators of the form
\begin{equation}
C_\Gamma(\vec k,t)
\equiv\sum_{\vec x}
\langle \bar\psi\Gamma\psi(\vec0,0)\;\bar\psi\Gamma\psi(\vec x,t)\rangle
e^{i\vec k.\vec x},
\end{equation}
which carry
zero baryon number. In [5] we showed that the generic
large-distance behaviour
in any given channel is  dominated by zero-energy particle-hole pairs, and as a
result the decay is algebraic, ie. $C_\Gamma(t)\propto t^{-\lambda(\vec k)}$,
with the
exponent $\lambda$ determined by the geometry of the overlap between two Fermi
disks with relative displacement $\vec k$ between their centres.
A particularly favourable configuration occurs for $\vert\vec
k\vert\approx2\mu$, in which case the Fermi surfaces just kiss,
and $\lambda={3\over2}$. In this paper we focus on point-split
meson operators corresponding to spatial components of the conserved vector
current, ie. of the form
\begin{equation}
j_i(x)={\eta_i(x)\over2}\left[\bar\chi(x)\chi(x+\hat\imath)+\bar\chi(x+\hat
\imath)\chi(x)\right],
\end{equation}
where $\chi,\bar\chi$ are staggered fermion fields, $\eta_1(x)=(-1)^{t}$,
and $\eta_2(x)=(-1)^{t+x_1}$. We set $\vec k=(k,0)$,
and define $C_\parallel$ in terms of the correlator $\langle
j_1(0)j_1(x)\rangle$ and $C_\perp\sim\langle j_2(0)j_2(x)\rangle$. In
Fig.~\ref{fig:results} we show $\vert C_{\parallel,\perp}\vert(k,t)$
data taken on a $48^3$ lattice
at $\mu=0.5$, for both
$k={\pi\over3}\approx2\mu$ and $k=\pi-{\pi\over3}$ (the modulus is taken
because we use a log-linear scale, and $C$ fluctuates in sign). Whereas
$C_\perp$ and $C_\parallel(k={\pi\over3})$ all show behaviour consistent with
algebraic
decay, the decay in the $C_\parallel(k=\pi-{\pi\over3})$ channel is much
faster, and resembles the exponential decay expected of an isolated pole.
We have fitted it with the form
\begin{equation}
C_\parallel(\pi-k,t)=A\exp(-\Omega(k)t)+(-1)^tB\exp(-\omega(k)t),
\label{eq:zsfit}
\end{equation}
in most cases employing datapoints with $t\in[10,38]$.
For small $k$, the coefficient $B\gg A$, suggesting that the correlator is
dominated by a pole in the alternating channel. In Fig.~\ref{fig:results}
we plot the resulting dispersion relation $\omega(k)$ for $\mu=0.2,\ldots,0.6$.
The behaviour $\omega\propto k$ as $k\to0$ suggests the presence of a massless
pole similar to that of a phonon. Support for this interpretation
comes from recasting the meson bilinear in terms of continuum-like fields
$q^{\alpha a}$, $\bar q^{\alpha a}$ having spinor index $\alpha=1,\ldots, 4$ and
`color' index $a=1,2$ [6]. We obtain
\begin{equation}
(-1)^{x_1}(-1)^t\bar\chi(x)\eta_1\chi(x+\hat1)\sim i\bar
q(\gamma_0\otimes\tau_2^*)q,
\end{equation}
demonstrating that the excitation can be viewed
as an oscillation of local baryon density.

Although we are examining the $k\to0$ limit, we are trying to model a
Fermi
surface phenomenon on the lattice;
the appropriate scale with which to compare $\beta_0$
is the ``bare'' Fermi velocity (ie. with no discretisation
correction) which for free
fields with $E(k)=-\mu+\sinh^{-1}(\sin k)$ is given by
\begin{equation}
\beta_F^{\rm bare}={{\partial E}\over{\partial k}}\biggr\vert_{E=0}=
\sqrt{{1-\sinh^2\mu}\over{1+\sinh^2\mu}}.
\end{equation}
The general trend  in our data is that the speed ratio $s=\beta_0/\beta_F^{\rm bare}$ increases towards unity
as $\mu$ increases (for $\mu=0.2$ $s=0.896(8)$ and for $\mu=0.6$ $s=0.959(2)$).
In [3] we used Fermi liquid interaction to leading non-trivial order in 
$1/N_f$ to find analytical solutions to the Boltzmann equation corresponding to zero sound. The excitation 
is spin and isospin symmetric, and has speed ratio $s>1$ for almost all $\mu>\mu_c$. To what extent can we be sure that 
our numerical results describe the same physical phenomenon? 
Solutions with $s<1$ allow the possibility of emission of a phonon with momentum 
$\vec{q}$ and energy $\beta_0 |\vec{q}|$ from a quasiparticle with momentum $\vec{k}$ 
and energy $\mu+\beta_F(\vert\vec k\vert-k_F)$.
This is allowed kinematically for $s<1$, the angle of
emission $\phi$ satisfying
\begin{equation}
\cos\phi=s+{{\vert\vec q\vert}\over{2\vert\vec k\vert}}(1-s^2)>s.
\end{equation}
All radiation is emitted within a cone of half-angle $\cos^{-1}s$ centred on
the quasiparticle trajectory. This well-known phenomenon is variously known
as Landau damping, \v Cerenkov radiation, or most
appropriately in the current context, as a sonic boom.

First let us argue how our numerical results can support
a state resembling a simple
pole but with $s<1$, in apparent contradiction to the above.
On a spacetime lattice, the radiation process is constrained because there is a
natural lower bound for the angle of emission, $\phi_{\rm
min}\sim2\pi/L_s\vert\vec q\vert$. Landau damping is thus kinematically
forbidden for
\begin{equation}
s>\cos\phi_{\rm min}\simeq1-{{2\pi}\over{L_s^2\vert\vec q\vert^2}}.
\end{equation}
With a conservatively high $\vert\vec q\vert\sim\mu/2$, then on
a $48^3$ lattice we have no damping for $s>0.73$ at $\mu=0.2$, rising to
$s>0.97$ at $\mu=0.6$. It is plausible therefore that
Landau damping is suppressed in our lattice data and that the phonon
is described by an isolated pole even if $s<1$.

Simulations on volumes
considerably larger than those used here, however, will be needed to disentangle the
various possible systematic effects due to finite $L_s$, finite $\mu/T$, and
non-zero lattice spacing in order to determine the sign of $s-1$ for finite $N_f$.
We think the most probable explanation for $s<1$ is an effect of non-zero
$T\equiv(aL_t)^{-1}$;
the physical temperature of the $L_t=48$
lattice decreases by a factor of roughly two
going from $g^{-2}=0.75$ to $g^{-2}=0.6$, and
according to our lattice data $\vert s-1\vert$ decreases by
roughly the same factor. 
It is known empirically that the speed of sound in liquid $^3\mbox{He}$ {\em
increases} as $T\to0$ and the dominant mode of propagation changes from first
sound to zero sound [7].
In principle, since our extracted value of $s\lapprox1$
has relied on a rescaling
of $\beta_F$ to take account of discretisation artifacts, we
also need to go to considerably
finer lattices, ie with $\beta_F^{\rm bare}\approx1$,
in order to demonstrate a clear distinction
between our signal and first sound with expected
propagation speed $\beta_1\simeq0.7$.
If, however, the theoretical arguments about zero sound dominating as $T\to0$
can be taken seriously, then for the
first time we have succeeded
in identifying a collective oscillation in a lattice
simulation.


\begin{thebibliography}{0}

\bibitem{Simon}
S.J.~Hands,
Nucl.\ Phys.\ A {\bf 642} (1998) 228.

\bibitem{landau57}
L.D. ~Landau, 
Sov. Phys. JETP {\bf 5} (1957) 101.

\bibitem{HS}
S.J.~Hands and C.G.~Strouthos, {\tt hep-lat/0406018}.

\bibitem{HM}
S.J.~Hands and S.E.~Morrison,
Phys.\ Rev.\ D {\bf 59} (1999) 116002.

\bibitem{HKTS}
S.J.~Hands, J.B.~Kogut, C.G.~Strouthos and T.N.~Tran,
Phys.\ Rev.\ D {\bf 68} (2003) 016005.


\bibitem{BB}
C.~Burden and A.N.~Burkitt, Europhys.\ Lett.\  {\bf 3} (1987) 545.

\bibitem{AGDNO}
J.W. Negele and H. Orland, {\sl Quantum Many-Particle Systems\/},
(Westview Press, Boulder, 1988) ch. 6.

\end{thebibliography}
\end{document}